# Dependency of U.S. Hurricane Economic Loss on Maximum Wind Speed and Storm Size

Alice R. Zhai

La Cañada High School, 4463 Oak Grove Drive, La Canada, CA 91011

Jonathan H. Jiang

Jet Propulsion Laboratory, California Institute of Technology, Pasadena, CA, 91109

Corresponding Email: Jonathan.H.Jiang@jpl.nasa.gov

**Abstract**: Many empirical hurricane economic loss models consider only wind speed and neglect storm size. These models may be inadequate in accurately predicting the losses of super-sized storms, such as Hurricane Sandy in 2012. In this study, we examined the dependencies of normalized U.S. hurricane loss on both wind speed and storm size for 73 tropical cyclones that made landfall in the U.S. from 1988 to 2012. A multi-variate least squares regression is used to construct a hurricane loss model using both wind speed and size as predictors. Using maximum wind speed and size together captures more variance of losses than using wind speed or size alone. It is found that normalized hurricane loss ($L$) approximately follows a power law relation with maximum wind speed ($V_{max}$) and size ($R$). Assuming $L=10^c V_{max}^a R^b$, $c$ being a scaling factor, the coefficients, $a$ and $b$, generally range between 4-12 and 2-4, respectively. Both $a$ and $b$ tend to increase with stronger wind speed. For large losses, a weighted regression model, with a being 4.28 and b being 2.52, produces a reasonable fitting to the actual losses. Hurricane Sandy's size was about 3.4 times of the average size of the 73 storms analyzed. Based on the weighted regression model, Hurricane Sandy's loss would be 21 times smaller if its size were of the average size with maximum wind speed unchanged. It is important to revise conventional empirical hurricane loss models to include both maximum wind speed and size as predictors.

1. Introduction

Landfalling hurricanes cause large amounts of economic damage, injury and loss of life. In the United States, hurricane losses account for the largest fraction of insured losses from all natural hazards (Bevere et al. 2013). It has been expected that hurricane loss (*L*) would depend on maximum wind speed ($V_{max}$), storm size (*R*), precipitation, storm surge, duration, path, and other factors such as building structure and population density (Vickery et al., 2006). Many studies have examined the dependency of hurricane loss on maximum wind speed and applied the wind speed dependency in empirical loss models. However, the relationship between hurricane loss and size has not been quantified, partly due to the lack of accurate measurements of storm size for the past historical events.

The dependency of hurricane loss on maximum wind speed follows an approximate power-law relationship, i.e., $L = \alpha V_{max}^{\beta}$, with *β* ranging between 3 and 9 (Pielke 2007; Nordhaus, 2010; Bouwer and Botzen 2011; Emanuel 2011) and *α* being a scaling factor, both depending on different cases or studies. Murnane and Elsner (2012) analyzed normalized US hurricane losses from 1900 to 2011 by a linear fitting between $\log_{10}(L)$ and wind speed for the top 10%, 25%, 50%, 75%, and 90% of hurricane losses. This method, called quantile regression approach, suggested an exponential relationship between normalized loss and wind speed.

Emanuel (2005) provided a theoretical basis for a possible relation between hurricane loss and its size. He expressed the total power dissipation (*PD*) of a tropical cyclone (TC) as the integral of the cubic of maximum wind speed over the size (*R*) of a storm through its lifetime (*τ*)

$$PD = 2\pi \int_0^\tau \int_0^R C_D \rho |V|^3 r\, dr\, dt, \qquad (1)$$

where $C_D$ is the surface drag coefficient, *ρ* is the surface air density, |*V*| is the magnitude of surface wind speed, *r* is the radius of the storm, and the integral is from storm center to the outer storm limit and over the lifetime(*τ*) of the storm. Since the economic loss of a hurricane is

driven by PD, Equation (1) shows that loss would increase with the squares of average storm size. However, due to the lack of historical data for size, Emanuel (2005) simplified the Eq (1) to omit the size dependency in actual calculations of *PD*.

Hurricane Sandy in 2012 reminded us that storm size should not be ignored when considering hurricane loss. A unique characteristic of Sandy is its enormous size when the tropical cyclone merged with a mid-latitude frontal system. At its peak size, Sandy's tropical storm-force winds (wind speed greater than 34 knots) spanned 1,100 miles, making it the largest Atlantic hurricane on record in terms of size. Hurricane Sandy made landfall on October 29, 2012 as a Category 1 hurricane, and its storm surge devastated coastal New Jersey and New York. The normalized loss for Sandy is 51.2 billion in 2013 US dollars based on the ICAT Damage Estimates (http://www.icatdamageestimator.com), making Sandy the eighth most costly storm since 1900. Out of the top 10 most expensive storms, Sandy is the only Category 1 hurricane at landfall; all other storms are Category 3 and higher. Therefore, it is important to consider not only wind speed but also storm size to account for hurricane losses.

The objective of this study is to quantify the relationship between hurricane loss and the hurricane maximum wind speed and size using multi-variate regression method and provide an empirical model for hurricane loss using both maximum wind speed and size as predictors. The estimated hurricane losses by the bivariate regression model are compared with those from the simple regression models using maximum wind speed or size alone. In particular, the relative roles of maximum wind speed and size in determining Hurricane Sandy's loss are analyzed.

The structure of the paper is as follows. Section 2 describes the data for hurricane loss, maximum wind speed and size, as well as the analysis method. Sections 3 and 4 show the relationship between loss and maximum wind speed, and between loss and size, respectively. Section 5 presents the bi-variate regression results and the sensitivity of the regression coefficients. Section 6 discusses the importance of the size for Hurricane Sandy's loss, and conclusions are given in Section 7.

2. **Data and Approach**

The US hurricane loss data are downloaded from the ICAT Damage Estimator website (http://www.icatdamageestimator.com/viewdata). ICAT is an insurance company that provides catastrophe insurance coverage to business and homeowners in the US. The losses are normalized to 2013 US Dollars, taking into account of inflation, wealth and population differences between the years that landfalling hurricanes occurred (Pielke et al., 2008). The loss data include hurricanes since 1900. The maximum wind speed at landfall for each storm is also provided by ICAT.

For storm size, five metrics are available in the Tropical Cyclone Extended Best Track database maintained by the National Hurricane Center (NHC), dating back to 1988. *R34*, *R50* and *R64* represent the radii of a storm where wind speeds at 10 meter height above the surface are 34, 50 and 64 knots, respectively. $R_{max}$ represents the radius of maximum wind speed. $R_{out}$ is the radius of the outmost closed isobar, i.e., the outer limit of a storm. The size data are given at four quadrants for each storm at 6-hourly interval. Averages of radii at the four quadrants are used in this study, although different weights for each quadrant may be explored in the future. While *R50, R60* and $R_{out}$ are highly correlated with R34 with correlation coefficients close to 0.8, $R_{max}$ and *R34* are only weakly correlated with a correlative coefficient of 0.13 for all available size data since 1988. The correlation between normalized hurricane loss and $R_{max}$ is found to be less than 0.1. Therefore, only *R34* is used as a size metric for the regression models for loss. A total of 73 tropical cyclones that made landfall in the US between 1988 and 2012 form the basis of this analysis. Table 1 lists the 73 cases with storm name, date of landfall, normalized loss, maximum wind speed and *R34* in descending order of loss values.

To construct a best-fit model to hurricane loss, the multi-variate regression tool available in Microsoft Excel Data Analysis package is used. Loss is expressed as a function of maximum wind speed, a function of size, and a function of both wind speed and size. Sensitivity of the fittings to storm intensity (i.e., wind speed) is examined.

The regression tool yields $R^2$ as explained variance and p-value for statistical significance of each fit. The explained variance indicates how much variance of the predictant (y) can be accounted for by the regression model using the predictor(s). The higher $R^2$ corresponds to a better fitting in terms of capturing the variations of a predictant. The p-value is the probability of the fitting coefficients for each predictor being zero. In other words, it is the chance of the dependency of the predictant on a predictor being purely random. To reject the null hypothesis that the dependency is random at a 95% statistical significance level, the p-value should be less than 0.05. The smaller p-value is, it is more statically significant that the fitting coefficients are nonzero.

## 3. The relationship between loss and maximum wind speed

Figure 1 is a scatter plot between losses and maximum wind speeds for the 73 cases. Both quantities are expressed in logarithms of base 10. There is an approximate linear relation between loss and wind speed in logarithmic scale, suggesting a power-law relationship between $L$ and $V_{max}$. A least-squares linear fit yields $L = 10^{-1.39} V_{max}^{5.27}$, which gives an $R^2$ of 0.39. The economic loss model thus explains 39% of the variance of the loss with a p-value of $3.32e^{-9}$ for statistical significance level of 95%. This small p-value suggests that it is statistically significant at 95% level to reject a null hypothesis that the coefficient for $V_{max}$ equals zero. The correlation between the logarithms of $L$ and $V_{max}$ is 0.63. The calculated root-mean-square (RMS) for the least-squares fit residuals of $log_{10}(L)$ is 0.93. The RMS accounts for how accurate the model is when estimating the actual loss. A low RMS means the model's estimated values are close to the actual values while a high RMS means the model's estimated values are far off from the actual values. Therefore, a low RMS is preferred. Here, a RMS of 0.93 suggests that the fitting errors for losses are on average within a factor of 10.

## 4. The relationship between loss and storm size

Figure 3 shows the relationship between loss and storm size (R), represented by *R34*. Their logarithms exhibit an approximately linear relation, but with more scatter than the counterpart for loss and wind speed. The lead-squares fit yields $L = 10^{3.94} R^{2.36}$. This linear fit captures only 26% of the variance of the loss, with the corresponding correlation of 0.51 and a p-value of $5.04e^{-6}$ for statistical significance level of 95%. The RMS for the least-squares fit residue of $log_{10}(L)$ is 1.03.

## 5. Dependency of loss on maximum wind speed and size

Using multi-variate linear regression, a loss model using both wind speed and size as predictors can be obtained. The correlation between wind speed and size is about 0.3 for the 73 tropical cyclones, indicating that they could serve as two nearly "independent" variables for a prediction of losses.

Following the approximate power-law relations shown in preceding sections, a general form of the loss model is assumed to be

$$L = 10^c V_{max}^a R^b \qquad (2),$$

where *c* is a scaling factor represented by the regression y-intercept, while *a* and *b* are the regression coefficients (slopes) for maximum wind speed and size, respectively. Such coefficients, termed "elasticity" as in Nordhaus (2010), are found to be different for subsets of the data grouped by maximum wind speed, shown in Table 2. For all 73 tropical cyclone cases ($V_{max} \geq$ 35 mph), *a* is 4.18, *b* is 1.25, and *c* is -1.83. For Category 1 or higher hurricanes ($V_{max} \geq$ 75 mph), *a* is 4.98, *b* is 2.66, and *c* is -6.22. For major hurricanes of Category 3 or higher ($V_{max} \geq$ 110 mph), *a* and *b* increase to 11.97 and 4.44, respectively, and *c* is -24.62. When wind speed is greater than 120 mph, *a* and *b* slightly decrease to 9.97 and 3.52, respectively; however, the sample size is very small (only 8), the results may not be robust at these extremely high wind speeds. Figure 3 shows the general increasing trend of "elasticity" for wind speed and size with storm intensity. The higher elasticity on wind speed for stronger storms is consistent with the previous studies (Nordhaus, 2010; Murnane and Elsner, 2012) despite that fewer but more recent samples are examined here. The increased power-law dependency of loss on size for Category 1 and higher hurricanes suggests that it is particularly important to consider the impact of size on loss for high-intensity storms, which are generally associated with greater losses than weaker tropical storms.

Furthermore, the explained variances by the bi-variate regressions are noticeably higher when maximum wind speed is higher. For example, the value of $R^2$ increases from 43% for storms with $Vmax \geq 35$ mph to 69% for hurricanes with $Vmax \geq 75$ mph, as shown in Table 2, Column 3. This suggests that wind speed and size play greater roles in determining losses when storm intensity reaches a certain threshold value, for example, 75 mph. Other factors, such as storm path, duration, and building structure could mask the relationship of loss with wind speed and size when storm intensity is weak.

Table 2 also lists the regression coefficients and explained variances if only wind speed or size is used for the least-squares fit for each subset of samples (Columns 7 to 10 in Table 2). Using two predictors consistently captures more variance of losses than using either wind speed or size alone in any subsets of samples. For storms with maximum wind speed below 100 mph, using wind speed or size alone would yield a higher elasticity on wind speed or size than using wind speed and size together, while the opposite occurs when maximum wind speed is at or greater than 100 mph. However, the single-variate regression captures less than one-third of the total variance of losses when $V_{max} \geq 100$ mph. The dependency on size is between the 2$^{nd}$ and 4$^{th}$ power, generally of lower order than the dependency on wind speed.

As the least squares regressions are performed on the logarithm of losses, the fitted values of losses are much lower than actual losses at large loss values. This occurs because taking the logarithm of the losses would reduce the impact of the more destructive hurricanes in regression fittings. To amplify the impacts of the large losses, we applied weights to the original data. This can be done by multiplying $log_{10}(L)$, $log_{10}(V_{max})$, and $log_{10}(R)$ by a weighting function $W$ and using the multi-variate regression tool to perform a tri-variate regression of $Wlog_{10}(L)$ using $Wlog_{10}(Vmax)$ and $Wlog_{10}(R)$ and $W$ as predictors. In this regression, we set the y-intercept to zero so that the regression coefficients for $Wlog_{10}(Vmax)$ and $Wlog_{10}(R)$ are $a$ and $b$, respectively, while the regression coefficient for $W$ gives the constant $c$, as in Eq (2), because

$$W log_{10}(L) = a\, W log_{10}(Vmax) + b\, W log_{10}(R) + c\, W. \tag{3}$$

A natural choice of weights is the values of losses. We also explored using the square roots of losses as weights. The resulting regression coefficients using the two different weights are shown in Table 3. For all samples and the subsamples of Category 1 and higher hurricanes, the weighted regressions yield higher order elasticities on both wind speed and size than the unweighted regressions (except for the elasticity on wind speed for square root of loss is used as weight). The RMS of fitting residuals to $log_{10}(L)$ are somewhat larger when the weights are used, but the overall fitting errors are on the same order as the unweighted models. The RMSs of residuals for the regression models using the subsets of storms with $V_{max} \geq 75$ mph are much smaller than those from the regression models using all samples. This is consistent with the greater explained variance by maximum wind speed and size for stronger storms than weaker storms.

Using the weighted regressions, the estimated losses for hurricanes with large losses were much closer to the actual values, and the low biases were significantly reduced, as shown in Figure 4.

In all the regression models, Hurricane Katrina is underestimated because its loss involved many unaccounted-for human impacts, such as the high vulnerability of low-rising residential areas, which are beyond the physical factors considered here.

## 6. Importance of the size on Hurricane's Sandy's Loss

In the case of Hurricane Sandy, its maximum wind speed at landfall was 75 mph and its size was 385 nautical miles (nm) at landfall. Its maximum wind speed is about 90% of the average (84 mph) out of the 73 storms but its size is about 3.4 times of the average (113 nm) (Figure 5).

Using the loss weighted regression model for all storms,

$$L = 10^{-3.77} V_{max}^{4.28} R^{2.52}, \qquad (4)$$

the fitted loss for Hurricane Sandy is about 59.0 billion, quite close to the actual loss estimated at 51.2 billion, while the unweighted regression models based on all samples significantly underestimate the loss (<10 billion) and the weighted model by the square root of loss yields an estimate of 38.0 billion (Figure 6). Excluding Sandy in the bi-variate fits yields very similar power-law dependencies and the fitted losses for Sandy vary slightly.

Figure 7 shows the predicted losses using the weighted regression model, Eq (2), for storms with different combinations of wind speed and size for an average storm and Hurricane Sandy. If Hurricane Sandy were of only the average size of 113 nm, its loss would have been about 21 (i.e., $3.4^{2.52}$) times smaller than the actual loss. Clearly, the enormous size of Hurricane Sandy plays a predominant role in the economic loss.

Using the regression coefficients shown in the last three rows in Table 3 derived from storms with $V_{max} \geq 75$ mph, the fitted loss would be $9.9 billion, $51.8 billion, and $36.7 billion, for

unweighted, weighted by loss and weighted by square root of loss, respectively. In all these models, the dependency of loss on size for Hurricane Sandy is even greater than what illustrated in Figure 7.

## 7. Conclusions

The US normalized hurricane losses are found to have an approximate power-law relation with maximum wind speed and size, indicated by the radius of tropical-storm force winds. The power-law order for maximum wind speed ranges from between 4-12, while the power-law order for size is approximately between 2 and 4. The high elasticity on wind speed is consistent with previous studies (Bouwer and Botzen 2011, Howard et al. 1972, Nordhaus 2010). This study, for the first time, presents a quantitative relationship between loss and size using historical data.

The dependency on the storm size is consistent with the expectation that the potential destructiveness of a storm is proportional to the area of the tropical-storm force winds (Emanuel 2005). The exact elasticity (the power-law order) is sensitive to the storm intensity – stronger storms have higher order power-law dependency on wind speed and size than the weaker storms, suggesting that it is especially important to take into account storm size when estimating losses for high-intensity hurricanes.

Storm size by itself does not account for a large fraction of the variance of hurricane losses. However, using wind speed and size together explains much more variance of losses than using the wind speed alone. Based on this study, conventional single-variate empirical models based on only maximum wind speed for hurricane loss should be revised to include both wind speed and size as predictors.

For Hurricane Sandy, its enormous size contributes predominantly to the economic loss. Out of the 73 tropical cyclones that were examined, Sandy's size was 3.4 times of the average storm size, corresponding to at least 21 times greater economic loss than that by an average sized storm at the same maximum wind speed. The huge loss by Hurricane Sandy is clearly a demonstration of the impact of storm size on damage.

Although many other factors could contribute to hurricane losses, the simple regression models using maximum wind speed and size as predictors are able to provide the first-order estimate of storm damages. The quantitative dependencies reported here provide useful guidance for developing more comprehensive loss models for hurricane damage research, insurance needs, and hazard preparations.

**Acknowledgements.** We thank Drs. Lixin Zeng, Hui Su and Chengxing Zhai for helpful discussions and detailed comments on the manuscript. We thank Dr. Longtao Wu for providing the storm size data. JHJ performs the work at the Jet Propulsion Laboratory, California Institute of Technology, under contract with NASA.

**Table 1**. The list of landfalling Atlantic tropical cyclones used in the study. The normalized losses are in billions of 2013 US Dollars. The maximum wind speeds ($V_{max}$) are in miles per hour (mph). The radii of 34 knots wind speed (R34) are in nautical miles (nm).

| Rank | Name | Date | Loss ($Billion) | $V_{max}$ (mph) | R34(nm) |
|---|---|---|---|---|---|
| 1 | Katrina | 8/29/2005 | 96.56 | 125 | 175 |
| 2 | Andrew | 8/24/1992 | 53.77 | 170 | 97.5 |
| 3 | Sandy | 10/29/2008 | 51.21 | 75 | 385 |
| 4 | Wilma | 10/23/2001 | 24.56 | 120 | 181.25 |
| 5 | Ike | 9/13/2008 | 20.32 | 110 | 168.75 |
| 6 | Ivan | 9/16/2004 | 18.39 | 120 | 175 |
| 7 | Charley | 8/13/2004 | 17.5 | 150 | 78.75 |
| 8 | Hugo | 9/21/1989 | 17 | 140 | 137.5 |
| 9 | Rita | 9/24/2005 | 11.92 | 115 | 145 |
| 10 | Frances | 9/ 5/2004 | 11.65 | 105 | 155 |
| 11 | Jeanne | 9/26/2004 | 8.94 | 120 | 145 |
| 12 | Allison | 6/ 5/2001 | 7.68 | 50 | 95 |
| 13 | Floyd | 9/16/1999 | 7.47 | 105 | 162.5 |
| 14 | Irene | 8/27/2011 | 7.33 | 75 | 182.5 |
| 15 | Fran | 9/ 5/1996 | 6.09 | 115 | 181.25 |
| 16 | Opal | 10/3/1991 | 5.93 | 115 | 181.25 |
| 17 | Isabel | 9/18/2003 | 4.76 | 105 | 237.5 |
| 18 | Gustav | 9/ 1/2008 | 4.53 | 105 | 180 |
| 19 | Bob | 8/19/1991 | 3.46 | 105 | 112.5 |
| 20 | Georges | 9/28/1998 | 2.85 | 105 | 121.25 |
| 21 | Dennis | 7/10/2005 | 2.66 | 120 | 153.75 |
| 22 | Isaac | 8/28/2012 | 2.41 | 80 | 152.5 |
| 23 | Andrew | 8/25/1992 | 2.24 | 115 | 125 |
| 24 | Gordon | 11/15/1990 | 1.64 | 50 | 120 |
| 25 | Irene | 10/14/1995 | 1.33 | 80 | 90 |
| 26 | Lili | 10/2/1998 | 1.27 | 90 | 142.5 |
| 27 | Bonnie | 8/26/1998 | 1.26 | 110 | 156.25 |
| 28 | Georges | 9/25/1998 | 1.19 | 105 | 112.5 |
| 29 | Dolly | 7/23/2008 | 1.11 | 85 | 110 |
| 30 | Alberto | 7/ 3/1994 | 1.03 | 65 | 60 |
| 31 | Erin | 8/ 2/1995 | 0.69 | 85 | 87.5 |
| 32 | Erin | 8/ 3/1995 | 0.69 | 100 | 87.5 |
| 33 | Fay | 8/19/2008 | 0.59 | 65 | 65 |
| 34 | Ernesto | 8/29/2006 | 0.56 | 45 | 31.25 |
| 35 | Bertha | 7/12/1996 | 0.51 | 105 | 143.75 |
| 36 | Josephine | 10/6/1992 | 0.5 | 70 | 50 |
| 37 | Isidore | 9/26/2002 | 0.49 | 65 | 235 |
| 38 | Cindy | 7/ 5/2005 | 0.38 | 75 | 37.5 |
| 39 | Gabrielle | 9/14/2001 | 0.35 | 70 | 87.5 |
| 40 | Marco | 10/10/1986 | 0.27 | 35 | 87.5 |
| 41 | Hermine | 9/ 6/2010 | 0.26 | 70 | 37.5 |
| 42 | Dennis | 9/ 4/1999 | 0.26 | 70 | 97.5 |
| 43 | Claudette | 7/15/2003 | 0.25 | 90 | 96.25 |
| 44 | Gilbert | 9/16/1988 | 0.25 | 70 | 237.5 |
| 45 | Chantal | 8/ 1/1989 | 0.24 | 80 | 100 |
| 46 | Danny | 7/19/1997 | 0.18 | 80 | 45 |
| 47 | Hanna | 9/ 6/2008 | 0.17 | 70 | 106.25 |
| 48 | Jerry | 10/14/1985 | 0.17 | 85 | 87.5 |
| 49 | Gaston | 8/29/2004 | 0.17 | 75 | 47.5 |
| 50 | Earl | 9/ 2/1998 | 0.14 | 80 | 83.75 |
| 51 | Mitch | 11/4/1994 | 0.14 | 65 | 143.75 |
| 52 | Charley | 8/14/2004 | 0.12 | 80 | 63.75 |
| 53 | Bret | 8/22/1999 | 0.1 | 115 | 67.5 |
| 54 | Charley | 8/22/1998 | 0.087 | 45 | 106.25 |

| 55 | Ophelia | 9/15/2005 | 0.083 | 75 | 97.5 |
| 56 | Bill | 6/30/2003 | 0.071 | 60 | 91.25 |
| 57 | Humberto | 9/13/2007 | 0.054 | 90 | 37.5 |
| 58 | Arlene | 6/20/1993 | 0.047 | 40 | 31.25 |
| 59 | Barry | 8/ 5/2001 | 0.046 | 70 | 60 |
| 60 | Hanna | 9/14/2002 | 0.03 | 60 | 62.5 |
| 61 | Harvey | 9/21/1999 | 0.025 | 60 | 43.75 |
| 62 | Earl | 9/ 2/2010 | 0.019 | 70 | 170 |
| 63 | Gordon | 9/17/2000 | 0.017 | 65 | 118.75 |
| 64 | Keith | 11/22/1984 | 0.015 | 65 | 162.5 |
| 65 | Alex | 6/30/2010 | 0.011 | 70 | 137.5 |
| 66 | Beryl | 8/ 9/1988 | 0.008 | 50 | 62.5 |
| 67 | Florence | 9/ 9/1988 | 0.007 | 80 | 75 |
| 68 | Kyle | 10/10/1998 | 0.007 | 40 | 10 |
| 69 | Fay | 9/ 7/2002 | 0.007 | 60 | 42.5 |
| 70 | Alex | 8/ 3/2004 | 0.005 | 80 | 63.75 |
| 71 | Chris | 8/28/1988 | 0.003 | 45 | 60 |
| 72 | Allison | 6/ 5/1995 | 0.003 | 70 | 87.5 |
| 73 | Gustav | 9/10/2002 | 0.0001 | 65 | 106.25 |

**Table 2. Regression results using maximum wind speed and/or size as predictors for loss, following the function form $L=10^c V_{max}^a R^b$. See text for details. $R^2$ is the explained variance of loss by a regression model.**

| Threshold $V_{max}$ | Sample Size | $R^2$ | a | b | c | $R^2$ ($V_{max}$ only) | a ($V_{max}$ only, b=0) | $R^2$ (R34 only) | b (R34 only, a=0) |
|---|---|---|---|---|---|---|---|---|---|
| >=35 | 73 | 0.45 | 4.19 | 1.25 | -1.83 | 0.39 | 5.27 | 0.26 | 2.36 |
| >=60 | 64 | 0.58 | 6.78 | 1.43 | -7.31 | 0.52 | 7.77 | 0.23 | 2.57 |
| >=65 | 60 | 0.55 | 6.92 | 1.44 | -7.62 | 0.48 | 7.69 | 0.18 | 2.32 |
| >=70 | 53 | 0.62 | 6.29 | 1.82 | -7.11 | 0.49 | 7.60 | 0.31 | 2.75 |
| >=75 | 43 | 0.69 | 4.98 | 2.66 | -6.22 | 0.40 | 7.11 | 0.51 | 3.36 |
| >=80 | 38 | 0.75 | 6.53 | 2.61 | -9.30 | 0.57 | 9.01 | 0.51 | 3.92 |
| >=85 | 30 | 0.75 | 6.82 | 2.48 | -9.64 | 0.50 | 8.07 | 0.41 | 3.10 |
| >=90 | 27 | 0.74 | 7.80 | 2.59 | -11.90 | 0.44 | 8.42 | 0.37 | 2.85 |
| >=100 | 24 | 0.64 | 8.82 | 3.13 | -15.17 | 0.30 | 6.73 | 0.16 | 2.09 |
| >=110 | 15 | 0.75 | 11.97 | 4.44 | -24.62 | 0.23 | 6.54 | 0.16 | 2.17 |
| >=115 | 13 | 0.80 | 12.11 | 4.34 | -24.72 | 0.25 | 6.92 | 0.20 | 2.31 |
| >=120 | 8 | 0.43 | 9.97 | 3.52 | -18.38 | 0.15 | 3.21 | 0.00 | -0.18 |

**Table 3. Regression results using maximum wind speed and size as predictors for loss, following the function form $L=10^c V_{max}^a R^b$ for unweighted and weighted fittings.**

|  | a | b | c | RMS of log10(L) |
|---|---|---|---|---|
| All storms ($V_{max}$ >= 35 mph), unweighted | 4.18 | 1.25 | -0.83 | 0.89 |
| All storms ($V_{max}$ >= 35 mph), weighted by loss | 4.28 | 2.52 | -3.77 | 1.21 |
| All storms ($V_{max}$ >= 35 mph), weighted by sqrt(loss) | 3.18 | 1.96 | -0.45 | 1.24 |
| Hurricanes ($V_{max}$ >= 75 mph), unweighted | 4.98 | 2.66 | -6.22 | 0.56 |
| Hurricanes ($V_{max}$ >= 75 mph), weighted by loss | 7.88 | 4.61 | -15.98 | 0.79 |
| Hurricanes ($V_{max}$ >= 75 mph), weighted by sqrt(loss) | 5.90 | 3.42 | -9.34 | 0.64 |

**Figure Captions**

**Figure 1:** The scatter plot of loss versus maximum wind speed for the 73 tropical cyclone cases. Both loss and wind speed are shown in logarithm at base 10.

**Figure 2.** The scatter plot of loss versus R34 for the 73 tropical cyclone cases. Both loss and R34 are shown in logarithm at base 10.

**Figure 3:** Variations of bi-variate regression coefficients, a for maximum wind speed and b size, with increasing threshold maximum wind speed, assuming that loss (L) follows the function form $L=10^c V_{max}^a R^b$.

**Figure 4:** The predicted losses using various regression models, compared to the actual losses for the 73 tropical cyclones cases. The inset is the zoomed-in figure for the top 10 large losses.

**Figure 5:** Hurricane Sandy's wind speed and size (in red) compared to the averaged wind speed and size (in blue) out of the 73 tropical cyclone cases.

**Figure 6:** The fitted losses (in blue) by several regression models for Hurricane Sandy, compared with the actual loss (in red).

**Figure 7:** Predicted losses for a storm of the average wind speed and size, or Hurricane Sandy's maximum wind speed and size. The loss-weighted bi-variate regression model is used to estimate the losses.

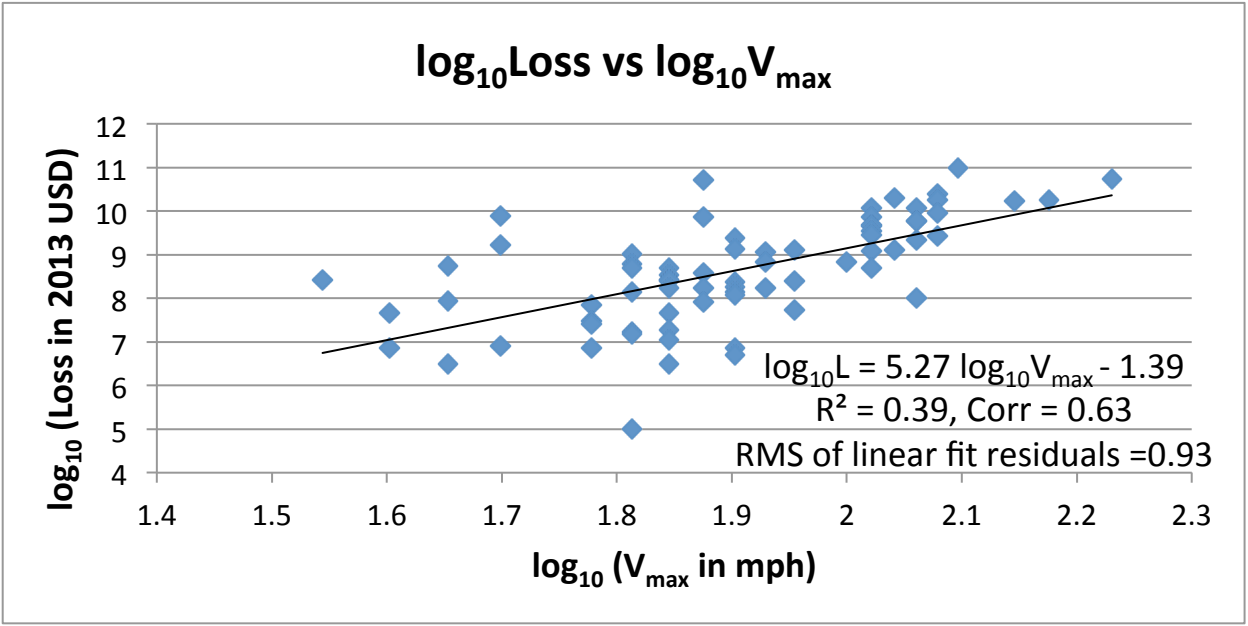

**Figure 1:** The scatter plot of losses versus maximum wind speed for the 73 tropical cyclone cases. Both losses and wind speeds are shown by their logarithm at base 10.

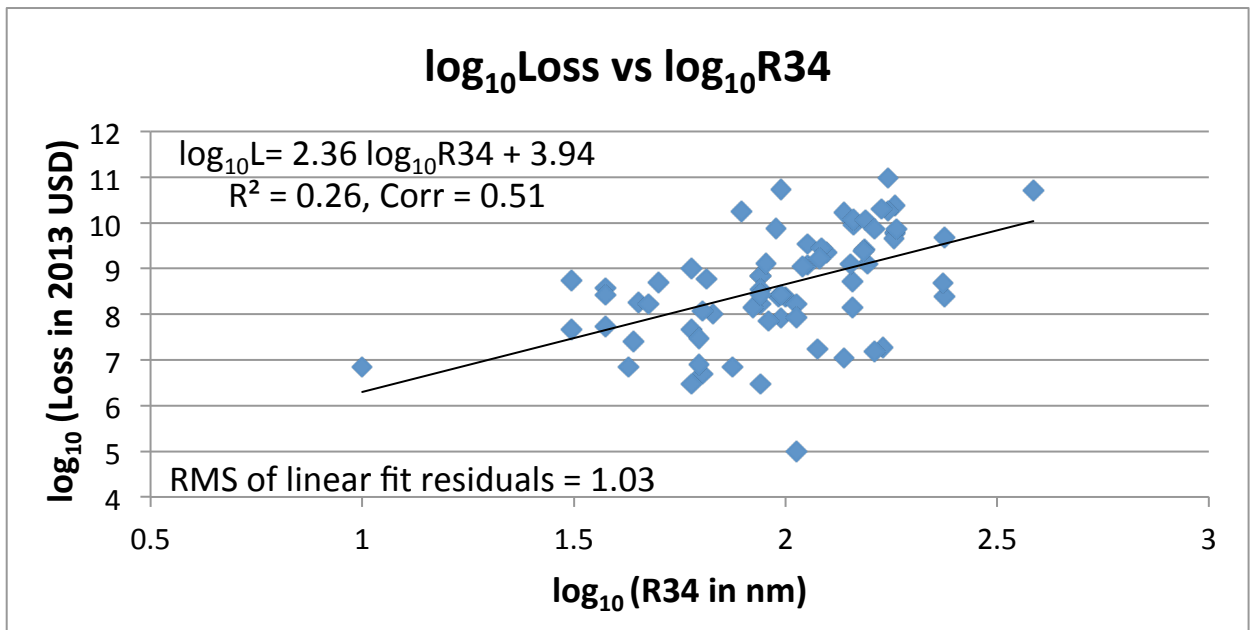

**Figure 2.** The scatter plot of losses versus R34 for the 73 tropical cyclone cases. Both losses and R34 are shown by their logarithm at base 10.

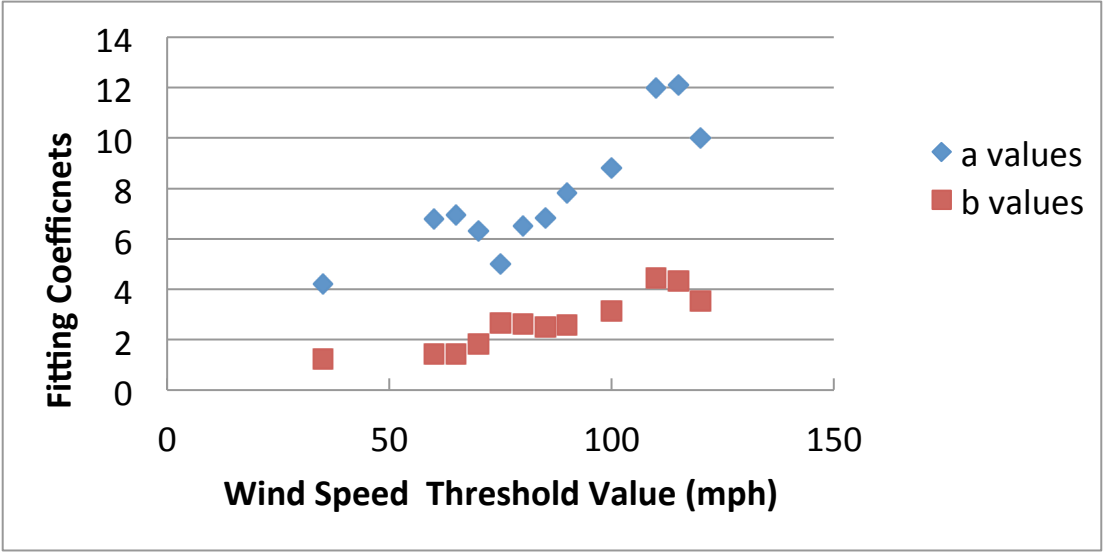

**Figure 3:** Variations of bi-variate regression coefficients, a for maximum wind speed and b size, with increasing threshold maximum wind speed, assuming that loss (L) follows the function form $L=10^c V_{max}^a R^b$.

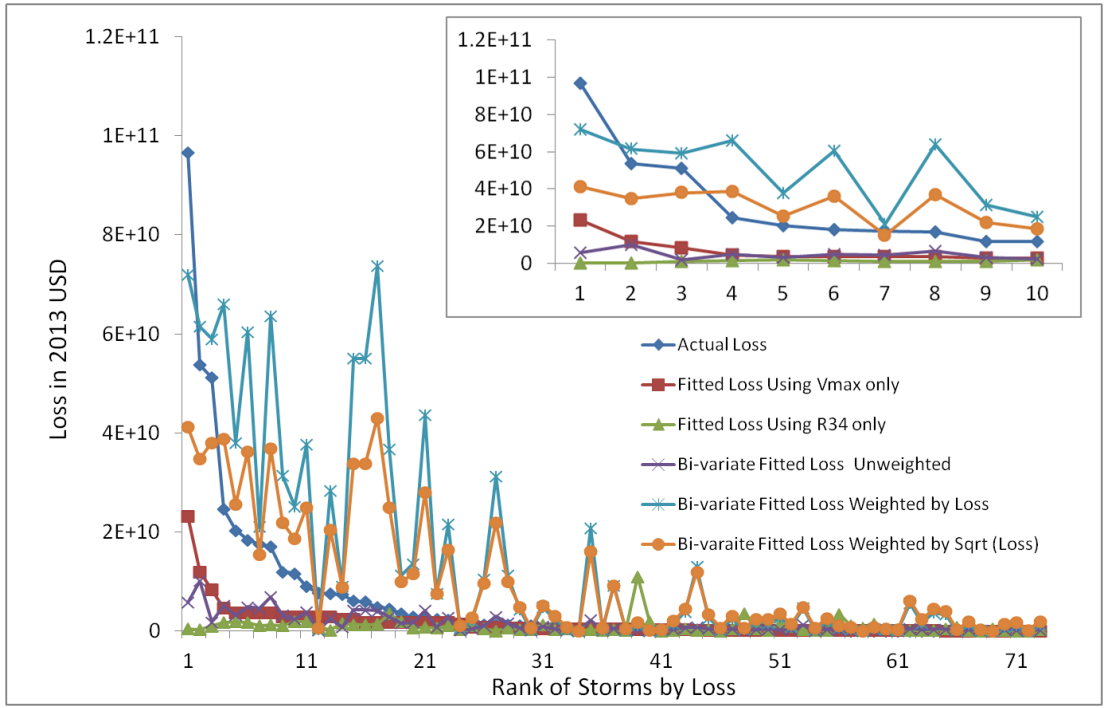

**Figure 4:** The predicted losses using various regression models, compared to the actual losses for the 73 tropical cyclones cases. The inset is the zoomed-in figure for the top 10 large losses.

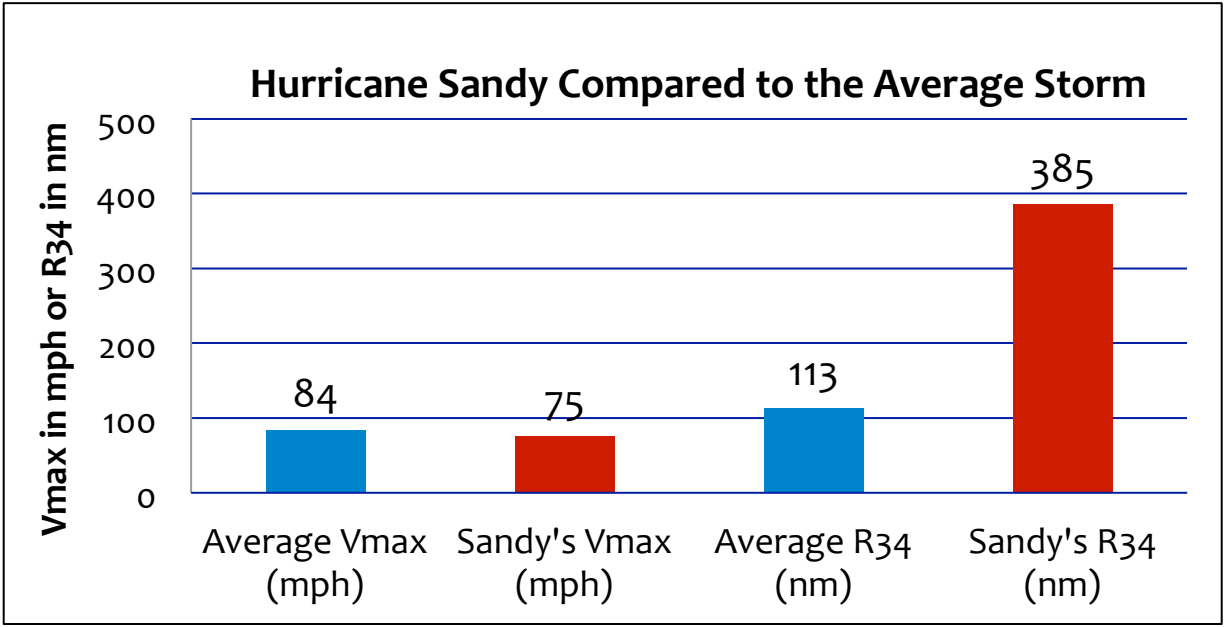

**Figure 5:** Hurricane Sandy's wind speed and size (in red) compared to the averaged wind speed and size (in blue) out of the 73 tropical cyclone cases.

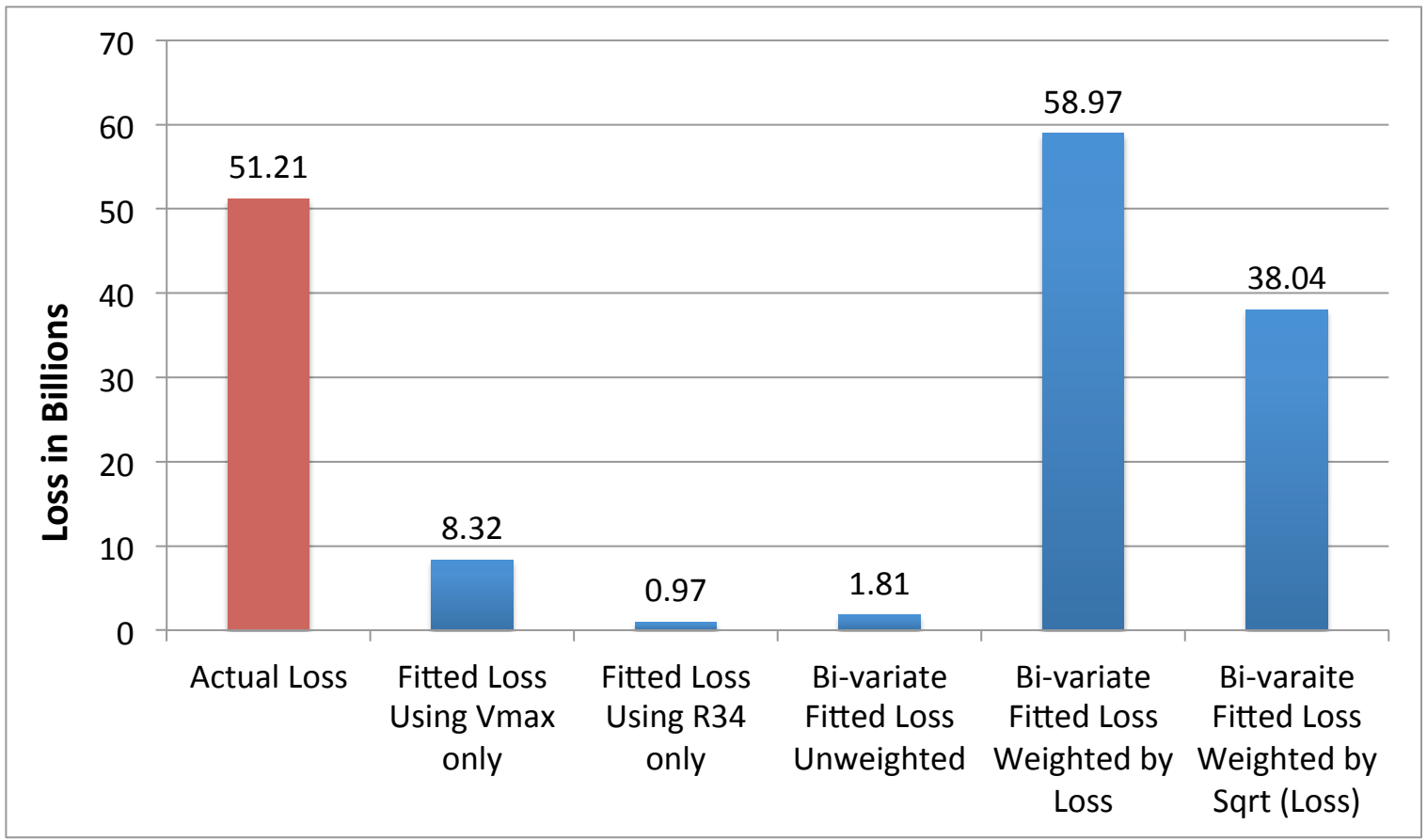

**Figure 6:** The fitted losses (in blue) by several regression models for Hurricane Sandy, compared with the actual loss (in red).

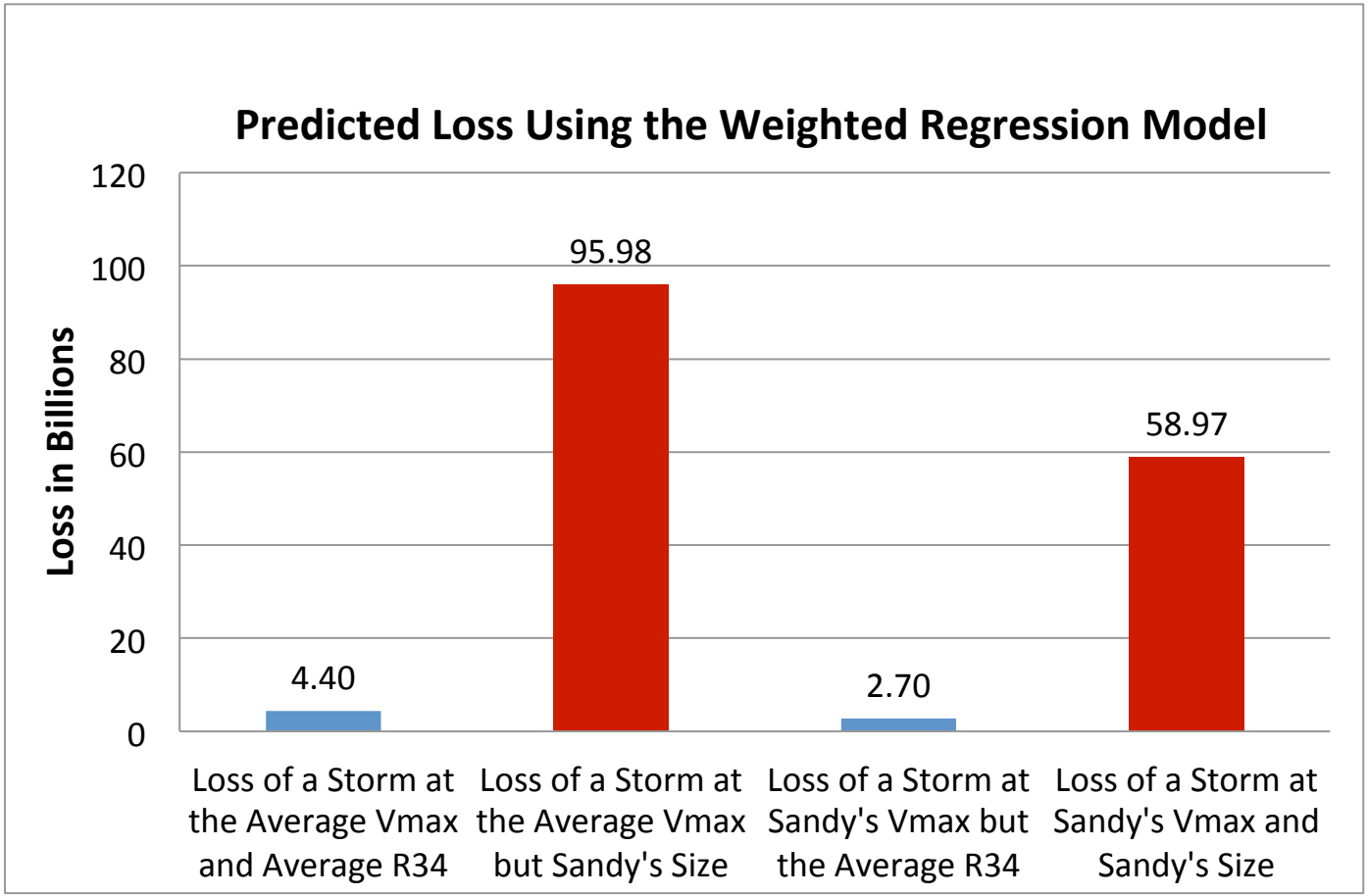

**Figure 7:** Predicted losses for a storm of the average wind speed and size, or Hurricane Sandy's maximum wind speed and size. The loss-weighted bi-variate regression model is used to estimate the losses.

24